\documentclass[prl,showpacs,twocolumn]{revtex4}

\usepackage{graphicx}

\begin{document}
\title{Quantum Localization in Open Chaotic Systems}
\author{Jung-Wan Ryu, G. Hur, and Sang Wook Kim}
\email{swkim0412@pusan.ac.kr}%
\affiliation{Department of Physics Education and Department of Physics,
Pusan National University, Busan 609-735, Korea}
\date{\today}

\begin{abstract}
We study a quasi-Floquet state of a $\delta$-kicked rotor with absorbing boundaries focusing on the nature of the dynamical localization in open quantum systems. The localization lengths $\xi$ of lossy quasi-Floquet states located near the absorbing boundaries decrease as they approach the boundary while the corresponding decay rates
$\Gamma$ are dramatically enhanced. We find the relation $\xi \sim \Gamma^{-1/2}$ and explain it based upon the finite time diffusion, which can also be applied to a random unitary operator model. We conjecture that this idea is valid for the system exhibiting both the diffusion in classical dynamics and the exponential localization in quantum mechanics.
\end{abstract}

\maketitle

Quantum localization (QL) is one of the fascinating phenomena which cannot be expected in classical mechanics. It has been observed in various
physical situations: Anderson localization (AL) in disordered systems \cite{And58}, dynamical localization (DL) in chaotic systems
\cite{Cas79}, weak localization in dirty metals \cite{Bergmann84}, and so on. The QL mainly originates from the interference among waves
returning their initial condition. Even though classical statistical mechanics predicts that a particle in a random potential exhibits stochastic
motion and thus gives rise to simple diffusion, the afore mentioned interference effect stops diffusion and
localizes it within some characteristic length scale referred to as the localization length, $\xi$. Due to the interference nature of the QL, the
coherence of the wave plays an indispensable role.

The DL is a dynamical version of the AL in the sense that the distribution in momentum space stops diffusing and exhibits typical exponential
localization. The DL was theoretically found in the quantum $\delta$-kicked rotor (DKR) \cite{Cas79,Izr90}, a paradigm of quantum chaos \cite{Reichl92}, and
experimentally realized by using ultracold atoms \cite{Moo95}. In the classical DKR the mean square deviation of momentum indefinitely increases
linearly in time, while in quantum mechanics it follows classical evolution only for a short time. At the characteristic time scale, namely {\em
a break time} it then begins to saturate so that eventually the quantum diffusion is completely suppressed. The formal equivalence between the DKR
and the Anderson model has been proved \cite{Fis82}.

In principle every {\em real} quantum systems are coupled to the environment since no information can be extracted from completely closed systems.
Thus it is a natural question to ask how the coupling of the quantum system to the environment modifies genuine quantum effects such as the QL.
Recently the interest in open quantum systems has been rapidly growing in quantum chaos community \cite{Casati97,Casati99,Ben01,Lu03,Sch04,Keating06}. In
the semiclassical limit the openness introduces two major modifications onto the chaotic quantum systems; (i) the fractal repellers manifest
themselves in quasi-eignstates \cite{Casati99} and (ii) the density of states follows the so-called {\em fractal} Weyl's law \cite{Lu03}. The
severe deviation from the random matrix theory has also been reported in chaotic scattering when the dwell time of an incident particle is
extremely short (shorter than the Ehrenfest time) \cite{Sch04}.

Even far from the semiclassical limit the openness of complicated quantum systems has been an interesting issue, for example, the
characteristics of lasing modes in chaotic microcavities \cite{Nockel97,Gma98,Lee02,Harayama03} (see \cite{Stone03,Kim07} for review) and localization of light in random media \cite{Wiersma97,Chabanov00,Sebbah06,Bliokh06}. In this Letter we investigate the {\em open} DKR in quantum mechanical
regime focusing on the characteristics of localization of lossy modes located near the open boundary. We found that quasi-eigenstates of the
open DKR are separated into two kinds; one is the localized state whose localization length is almost equivalent to that of the DKR {\em without}
the absorbing boundaries and the decay rate is determined from simple overlap argument discussed below. The other is a highly lossy mode located near the boundary, which is more strongly localized and of which decay rate is determined by considering the finite time classical diffusion. We also show that all these observation and explanation is applicable to an {\em open} random unitary operator model.

The Hamiltonian of the DKR is given as
\begin{equation}
\label{ham_kp} H = \frac{p^{2}}{2}+ k \sin x \sum_n \delta(t-nT),
\end{equation}
where $k$ is the kick strength, and $T$ is the time interval between successive kicks. We fix $k=14$ and $T=1$, i.e. $K=kT=14$ implying the
classical dynamics is fully chaotic. Note that the semiclassical limit implies $k\rightarrow \infty$ and $T\rightarrow 0$ with $kT$ kept
constant, so that we do not consider the semiclassical regime. The time evolution of the {\em open} quantum DKR with absorbing boundaries
\cite{Bor91} is described by $\left|\psi(T)\right>=\hat{P}\hat{U}\left|\psi(0)\right>$, where $\hat{U}$ is a unitary time evolution operator for one period without absorption, and the operator $\hat{P}$ projects the wavefunction to the states satisfying $|p| \leq p_c$, where $p_c$ represents the absorbing boundary. We choose $p_c=1000$ and set $\hbar=1$. This model has been extensively studied in the context of the fidelity decay or the Loschmit echo \cite{Casati99b}.

The DL manifests itself via the exponentially localized Floquet eigenstate of $\hat{U}$. In the {\em open} DKR the so-called {\em
quasi-}Floquet state (QFS) can be analogously defined as
\begin{equation}
\hat{P}\hat{U}\left|\phi\right> =
e^{-\Gamma}e^{i\alpha}\left|\phi\right>.
\end{equation}
Due to the openness of the system the Floquet eigenvalues no longer lie at a unit circle in a complex plane ($\Gamma > 0$). Each QFS can be
identified by its average momentum, $\bar{p} \equiv \left< \phi \left| \hat{p} \right| \phi \right>$. In one period $T$ the survival
probability of the given QFS is given as $e^{-2\Gamma}$ so that the life time of the state, $\tau_L$, is obtained by ${\tau_L}\propto
\Gamma^{-1} $.

\begin{figure}
\begin{center}
\includegraphics[width=0.48\textwidth]{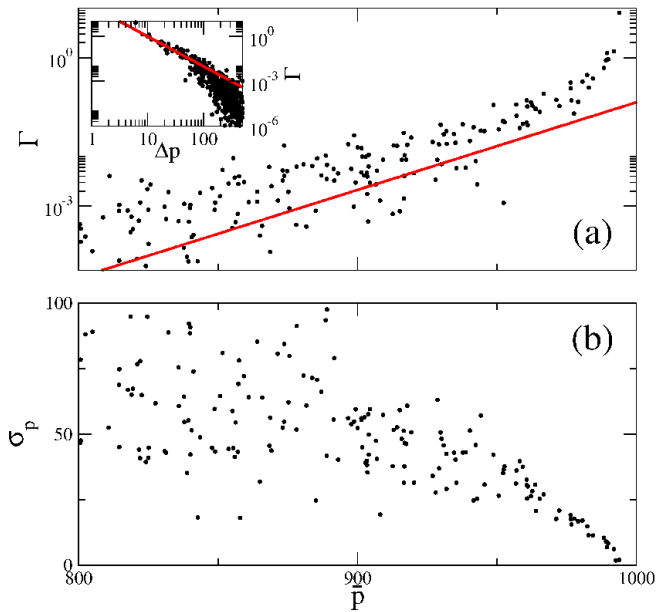}
\caption{(color online). (a) The decay constants $\Gamma$ of each QFS of the open DKR as a function of the average momentum $\bar{p}$ in log
scale. The straight line represents Eq.~(\ref{drate2}). The inset shows $\Gamma$ versus $\Delta p \equiv p_c - \bar{p}$ in log-log scale. The
straight line represents Eq.~(\ref{drate}). (b) The standard deviations $\sigma_p$ as a function of $\bar{p}$. The known theory for the DKR
without absorbing boundary predicts $\sigma_p \sim 50$.} \label{fig1}
\end{center}
\end{figure}

In Fig.~\ref{fig1} we present the decay constant $\Gamma$ and the width of the momentum distribution $\sigma_p$ of QFSs labelled by
$\bar p$. Here two visible features are clearly observed: (i) For
$\bar{p}$ much smaller than $p_c$, $\Gamma$ increases
exponentially and $\sigma_p$ remains constant with some
fluctuation. (ii) As $\bar{p}$ approaches $p_c$, $\Gamma$ more
rapidly increases and $\sigma_p$ linearly decreases.

\begin{figure}
\begin{center}
\includegraphics[width=0.48\textwidth]{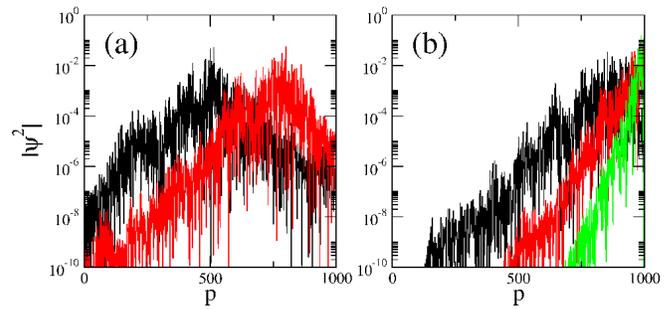}
\caption{(color online). (a) The momentum distributions of two typical exponentially localized QFS's with $\bar{p} \sim 502.86$ and $765.47$,
where $\tau_L > \tau_B$ is satisfied. (b) As $\bar{p}$ approaches $p_c$ through 865.80, 961.04, and 982.88 (from the left to the right), the distribution becomes narrower.}
\label{fig2}
\end{center}
\end{figure}

The first feature can be easily understood. For $\bar{p} \ll p_c$ the localized state has negligible influence from the absorbing boundary so
that the QFS is not so different from the original Floquet state of the DKR {\em without} absorbing boundaries as shown in Fig.~\ref{fig2}(a).
This is also true even quantitatively since it is fairly good to estimate $\sigma_p$ based upon the well-known relation $\sigma_p \sim \xi \sim
D/2 \sim K^2/4$ \cite{Shepelyanasky86}, where $D$ is the classical diffusion constant. It implies that the QFS can be described as $\phi \sim
\exp(-\left|p-\bar{p}\right|/\xi)$. The decay constant $\Gamma$ is then determined by considering the overlap between the exponential tail of
the localized QFS and the absorbing region given as $|p| \ge p_c$: $1 - e^{-2\Gamma} \sim \xi^{-1}\int^{\infty}_{p_c}
\exp(-2\left|p-\bar{p}\right|/\xi) dp$. It leads us to
\begin{equation}
\Gamma 
\sim 
\exp\left[-\frac{2}{\xi} \Delta
p\right], \label{drate2}
\end{equation}
where $\Delta p = p_c-\bar{p}$. Figure~\ref{fig1}(a) clearly shows
this expectation is correct.

As $\bar{p}$ approaches $p_c$, however, the absorbing boundary has a dramatic influence on the QFS's. They become much more lossy and
even more strongly localized, which is unlikely because the strongly localized mode has smaller overlap with the absorbing
region, i.e. becomes less lossy. It is worth mentioning that here we exploit the absorbing boundaries to open the system to the environment.
Usually the coupling to the environment introduces dissipation or decoherence, which destroys the coherence itself, let alone the
localization. Therefore, one expects that the localization length increases. In this sense the absorbing boundaries are special.

In a usual DKR there is only one important time scale, namely the break time $\tau_B$, at which the diffusion stops so as to
determine the localization of the Floquet state: ${\sigma_p}^2
\sim \xi^2 \sim D\tau_B$. In an {\em open} DKR, however, we should consider one
more time scale, the life time, $\tau_L$. A clue comes from the
fact that the crossover from constant to decreasing
$\sigma_p(\bar{p})$ takes place around $\tau_L \sim \tau_B$.

The system undergoes its meaningful dynamics only for $t<\tau_L$ in the sense that the probability distribution no longer changes except overall
decaying. If $\tau_L<\tau_B$ is satisfied, therefore, the meaningful dynamics stops before the break time is reached. It means that the classical diffusion plays a dominant role in the decay process since a particle disappears before the quantum suppression of classical diffusion takes place. In other words, the particle escapes when it diffusively arrives at the absorbing boundary. One can expect that the life time $\tau_L$ of a given QFS
with $\bar{p}$ is determined simply from the duration time for which the particle travels from $\bar{p}$ to $p_c$ by diffusion: $\tau_L \sim
(p_c-\bar{p})^2/D$. This can be rewritten as
\begin{equation}
\label{drate} \Gamma \sim \frac{D}{\Delta p^{2}},
\end{equation}
which works quite well as shown in the inset of Fig.~\ref{fig1}(a). It is emphasized that Eq.~(\ref{drate}) is non-trivial because we are not
dealing with semiclassical regime. The classical diffusive dynamics decides the decay constant even in the deep quantum regime when the loss is
large enough.

Considering the above argument the linear decrease of $\sigma_p$ can also be understood. The meaningful dynamics stops at $\tau_L$ ($<\tau_B$)
so that ${\sigma_p}^2$ is determined not from $D\tau_B$ but from $D\tau_L$. By using Eq.~(\ref{drate}) we obtain
\begin{equation}
\label{sigmax} \sigma_p \sim \sqrt{\frac{D}{\Gamma}} \sim \Delta
p,
\end{equation}
which is clearly seen in Fig.~\ref{fig1}(b).

In some sense very lossy modes near the absorbing boundary are not interesting since they decay so fast. They do not contribute to long time
dynamics so that they form only broad peaks even in the scattering cross section. Sometimes such a mode, nevertheless, becomes of great
importance; for example, a very lossy mode can play a dominant role in lasing operation, where an external energy input compensates the loss of
the mode \cite{Gma98}. It is also worth mentioning that the spatial shape of
an individual mode is recently measured in the experiment of light in a random media, where the lossy mode strongly localized near the boundary
is also observed \cite{Sebbah06,Bliokh06}.

The main theme of our work is that the lossy QFS of the open DKR near the absorbing boundary shows rather stronger localization. In principle this is
applicable to any system that exhibits diffusion in classical mechanics and exponential localization in quantum mechanics. We consider one more
example originating from one-dimensional Anderson model. Instead of Anderson's tight-binding Hamiltonian we exploit the so-called random unitary
operator $\hat{U} = \hat{D} \hat{S}$, where $\hat{D}_{mn}=e^{i\theta_m} \delta_{mn}$ with a random phase $\theta_m$ \cite{Kos91}, and an infinite dimensional matrix $\hat{S}$ is defined as
\begin{eqnarray}
\hat{S} & = & \left(\begin{array}{cccccc}
\ddots & rt & -t^{2}\\
 & r^{2} & -rt\\
 & rt & r^{2} & rt & -t^{2}\\
 & -t^{2} & -tr & r^{2} & -rt\\
 &  &  & rt & r^{2}\\
 &  &  & -t^{2} & -tr & \ddots\end{array}\right).
\end{eqnarray}
Here, $\hat{S}$ describes the scattering process via the relation $\phi'_i = \hat{S}_{ij}\phi_j$, where $\phi$($\phi'$) represents the incoming (outgoing) flux for a scatterer. In one-dimensional case two fluxes with the opposite direction at each site between two neighboring scatterers form two adjacent vector components, $\phi_{2k}$ and $\phi_{2k+1}$ \cite{Kos91}. The condition $r^{2}+t^{2}=1$ holds to ensure unitarity. For all calculation we fix $t=0.4$. In order to control the width of the off-diagonal components, which roughly corresponds to the parameter $K$ of a DKR, we simply
multiply $\hat{S}$: i.e. $\hat{U}_n = \hat{D} \hat{S}^n$ ($n=1,2,3,\cdots$). Note that the parameter $K$ of a classical DKR determines the
maximum momentum transfer to a particle from each kick. The non-zero $(i,j)$th off-diagonal component of $\hat{U}$ implies there exists a
non-zero transition probability between these two states: roughly speaking $K \sim \textrm{max}(i-j)$ satisfying
$\left<i\left|\hat{U}\right|j\right> \neq 0$. One then expects that the larger $n$ the bigger the corresponding {\em effective} $K$. We now introduce the projection operator $\hat{P}$ and the QFS's of $\hat{P}\hat{U}_n$ are investigated in a similar way.

\begin{figure}
\begin{center}
\includegraphics[width=0.48\textwidth]{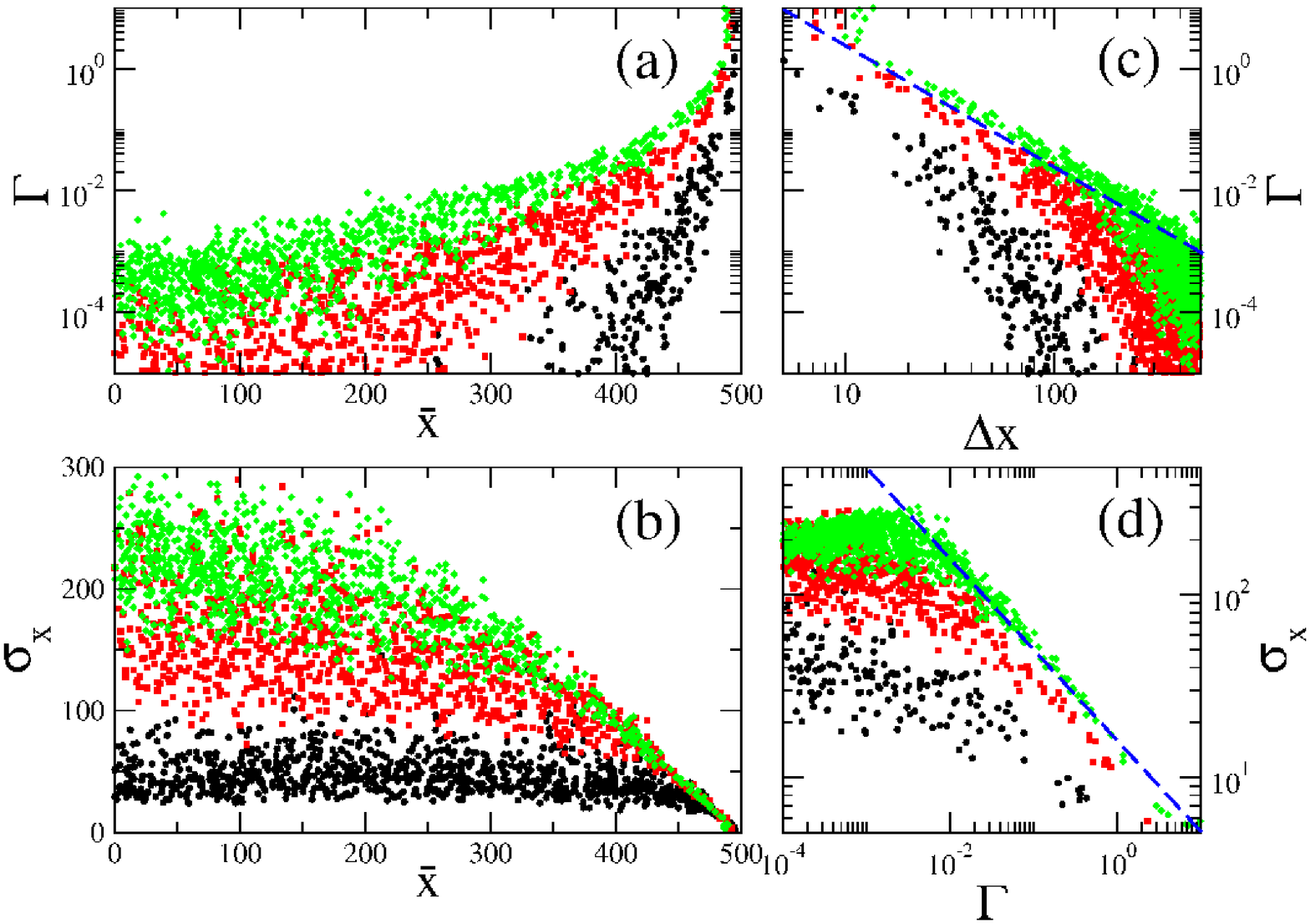}
\caption{(color online) From bottom to up $n=10$, 20, 30 are exploited for all figures. (a) and (b) are the same as Fig.~\ref{fig1}(a) and (b), respectively,
except that a random unitary operator model is considered. (c) shows $\Gamma$ versus $\Delta x $ in log-log scale. The straight line represents
Eq.~(\ref{drate}). (d) presents $\sigma_x$ versus $\Gamma$ in log-log scale. The straight line represents Eq.~(\ref{sigmax}).} \label{fig3}
\end{center}
\end{figure}

In Fig.~\ref{fig3}(a) and (b), we present the decay constant $\Gamma$ and the standard deviation $\sigma_x$, respectively, of an individual QFS of
$\hat{P}\hat{U}_n$ for various $n$. Once again two visible features are observed. For $\bar{x}$ ($\equiv \left< \phi \left| \hat{x} \right| \phi
\right>$) much smaller than $x_c$ ($x$ is used here instead of $p$), as $\bar{x}$ increases, $\Gamma$ exponentially increases and $\sigma_x$
remains constant. As $\bar{x}$ closely approach $x_c$, however, $\Gamma$ more rapidly increases and $\sigma$ decreases linearly. These
observation are exactly analogous to those obtained in the open DKR (see Fig.~\ref{fig1}). As $n$ increases, the crossover from constant to linearly
decreasing $\sigma_x$ is reduced since the bigger $n$ the larger the effective $K$, consequently the longer the break time. Figure \ref{fig3}
(c) and (d) reconfirms the relations (\ref{drate}) and (\ref{sigmax}), respectively, i.e. $\Gamma \propto \Delta x^{-2}$ and $\sigma \propto
\Gamma^{-1/2}$. It is emphasized that these results are independent of $n$.

\begin{figure}
\begin{center}
\includegraphics[width=0.48\textwidth]{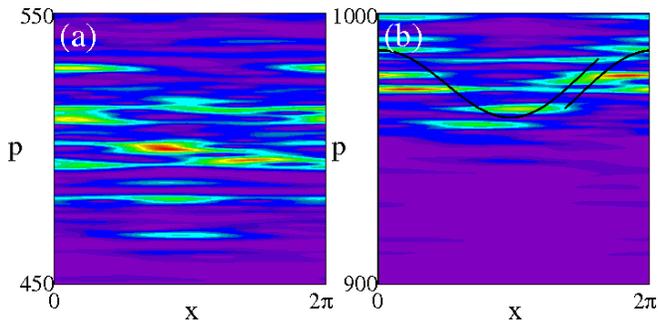}
\caption{(color online) Husimi distribution functions of the QFS for (a) $\bar{p} \simeq 502$ and (b) $\bar{p} \simeq 972$. The black curve represents the classical unstable manifold.}
\label{fig4}
\end{center}
\end{figure}

A final remark is in order. We have shown that the results obtained in the DKR can be directly applied to those of the random unitary operator
model. It is noted, however, that chaos is different from random motion. First, the chaotic dynamics has more structures in phase space, e.g.
stable and unstable manifolds associated with periodic orbits. In our DKR one expects they play no important role since we do not consider the
semiclassical limit. Indeed we deal with deep quantum regime. Nevertheless, the classical diffusion has a significant contribution to the decay
constant when the loss is large enough. Even in the deep quantum regime some reminiscences of
unstable manifolds or chaotic repeller manifest themselves in Husimi plots of the lossy modes as shown in Fig.~\ref{fig4}(b). However, the mode
far from the absorbing boundary does not show any similar underlying classical structure as shown in Fig.~\ref{fig4}(a). Note that such a structure does not exist in the random unitary operator model. Secondly, in chaotic system
there is one more important time scale, which is an Ehrenfest time $\tau_E$ defining classical to quantum crossover \cite{Sch04}. When $\tau_L < \tau_E$, the
characteristics of the corresponding QFS can be strongly modified \cite{Sch04}. In our case $\tau_E$ is extremely small so that one can safely
ignore it \cite{note2}.

In summary, we have shown that the QFS's with their life time smaller than the break time, i.e. $\tau_L < \tau_B$, exhibit rather stronger
localization and considerable loss. In this case, the main mechanism of the decay is determined from classical diffusion, which gives the
relation (\ref{drate}). In addition, before the break time is reached the classical diffusion effectively stops so that the state with much
narrower momentum distribution is formed. The width of the distribution is then described as Eq.~(\ref{sigmax}). Such a simple explanation can also be
successfully applied to a random unitary operator model. We believe that our theory is valid for various physical situation that the diffusion
takes place in classical mechanics and strong exponential localization exists in quantum mechanics. We hope that our expectation is
experimentally proven e.g. by direct observation of geometrical shapes of quasi-bound states in light in random media.

This work was supported by Korea Research Foundation Grant (KRF-2006-312-C00543) and by Korea Science and Engineering Foundation Grant
(R01-2005-000-10678-0).


\begin{thebibliography}{99}
\bibitem{note2}
The Ehrenfest time is given by $\tau_E=\ln N/\lambda$, where $\lambda$ is of order of the Lyapunov exponent determined as $\lambda \sim
\ln(K/2)$ in the DKR, and $N \sim 10^3$, so that one obtains $\tau_E \sim 1$.
\end{thebibliography}
\end{document}